# A DTCWT-SVD Based Video Watermarking resistant to frame rate conversion

Yifei Wang
School of Computer Science
Fudan University
Shanghai, China
20110240074@fudan.edu.cn

Qichao Ying
School of Computer Science
Fudan University
Shanghai, China
20110240050@fudan.edu.cn

Zhenxing Qian
School of Computer Science
Fudan University
Shanghai, China
zxqian@fudan.edu.cn

Sheng Li
School of Computer Science
Fudan University
Shanghai, China
lisheng@fudan.edu.cn

Xinpeng Zhang
School of Computer Science
Fudan University
Shanghai, China
zhangxinpeng@fudan.edu.cn

*Abstract*—Videos can be easily tampered, copied and redistributed by attackers for illegal and monetary usage. Such behaviors severely jeopardize the interest of content owners. Despite huge efforts made in digital video watermarking for copyright protection, typical distortions in video transmission including signal attacks, geometric attacks and temporal synchronization attacks can still easily erase the embedded signal. Among them, temporal synchronization attacks which include frame dropping, frame insertion and frame rate conversion is one of the most prevalent attacks. To address this issue, we present a new video watermarking based on joint Dual-Tree Cosine Wavelet Transformation (DTCWT) and Singular Value Decomposition (SVD), which is resistant to frame rate conversion. We first extract a set of candidate coefficient by applying SVD decomposition after DTCWT transform. Then, we simulate the watermark embedding by adjusting the shape of candidate coefficient. Finally, we perform group-level watermarking that includes moderate temporal redundancy to resist temporal desynchronization attacks. Extensive experimental results show that the proposed scheme is more resilient to temporal desynchronization attacks and performs better than the existing blind video watermarking schemes.

## I. INTRODUCTION

Social networks are clustered with shared video files due to the widespread of various display devices including cell phones, tablets and computers [1]. However, video content is increasingly vulnerable to tampering, copying and redistributing by attackers. For example, pirating videos can be a perilous means to jeopardize copyright protection and potentially make a fortune, which leaves the interests of owner of videos under threat. Hence, enhancing copyright protection via multimedia steganography and robust watermarking is within center research in the past decades.

Digital watermarking has emerged as an effective technology and a promising solution to identify the ownership of a digital product [2-3]. It can help effectively declaring the ownership, tracking the criminal chain and cutting off the illegal dissemination [15-16]. Digital video watermarking embeds digital copyright information covertly into the carrier video, and extract the information with high accuracy at the recipient's side [4-8]. Robustness, imperceptibility and capacity are three main evaluation metrics for digital watermarking. First, the embedding must not affect the normal usage of the cover video or be discovered by the third party. Second, the hidden data must survive multiple kinds of digital attacks including quantization, compression, rescaling, etc. Third, the payload needs to be large enough in order to hold information for identification and forensics use. Compared to image watermarking [17-19], video watermarking has two extra challenges. On the one hand, videos have an additional temporal dimension compared to images, and therefore watermarking or videos need to be resilient against both spatial and temporal synchronization attacks. On the other hand, human eyes are more sensitive to cross-frame perturbations compared to in-frame distortions, which poses a stricter restriction on the magnitude of video embedding to maintain the consistency of the watermarked frames with the neighboring frames.

Frame rate conversion attack changes the number of frames per second displayed in video. It is mostly done by periodically adding or removing frames to influence the frame-per-second (fps) rate. Such attacks are very common and trivial in our daily life. For example, when an online video with a low fps is played on a high fps enabled device, the cloud server might resample the video to increase the frame rate so that the video can be played more vividly. In the same fashion, the fps might be lowered if the network condition is poor. There are more cased that might include other kinds of desynchronization attacks such as frame averaging, frame swapping etc. These attacks can easily fail the process of watermark extraction where synchronization is the first and foremost step. Previously, there have been a number of robust video watermarking schemes for copyright protection [4-8]. While most of these schemes can ensure imperceptible embedding as well as robustness against various spatial attacks, none of these methods has properly and blindly addressed the issue poor robustness against temporal synchronization attack.

In this paper, we propose a robust video watermarking scheme, jointly against desynchronization attacks as well as common geometric attacks, including rotation, cropping and scaling. To combat video desynchronization attacks, our intuition is that the total duration of video remains unchanged after frame rate conversion, and therefore a proper division of video clip can effectively achieve synchronization of watermark in temporal dimension. To be specific, we adjust the shape of candidate coefficient in the DTCWT-SVD domain. Meanwhile, the duration of video is exploited in the scheme to create a group-level watermark for making it robust against frame rate conversion. The experimental results show that our scheme is robust against combined attacks of both geometric attacks and temporal synchronization attacks. Also, the imperceptibility of the embedding in our method is superior to the existing methods.

The rest of this paper is organized as follows. Section II discusses the related works on robust video watermarking and summarizes the preliminary work of this article. Section III introduces the method proposed in this paper in detail. Section IV shows the experimental results and comparisons. Section V concludes the paper.

## II. RELATED WORK

### A. Robust Video Watermarking

In the past decades, many video watermarking methods based on transform domains including DFT [4], DCT [5-6], DWT [7-8], DTCWT have been proposed by researchers to achieve the watermarking objectives. In [4], a class of generalized correlators is constructed based on the generalized Gaussian distributions. Wang et al. [5] proposes a set of robust MPEG-2 video watermarking techniques, focusing on commonly used typical geometric processing for bit-rate reduction, cropping, removal of any rows, arbitrary-ratio downscaling, and frame dropping. It is well notifying that while some innovative image watermarking schemes have been proposed that utilize deep networks for generalized robustness [20-21], there is much less work that apply deep learning to conduct video watermarking [22].

DTCWT-based video watermarking schemes benefit from the advantage of shift invariance, direction selectivity and perfect reconstruction. The magnitude of the lower-frequent coefficients remains nearly unchanged when the video frames go through a geometric distortion including rotation, scaling and cropping. DTCWT-based methods are with improved robustness to geometric attacks as compared to the other transform methods. For example, Lino et al.[10] proposes a DTCWT-based video watermarking scheme which is resilient to rotation, cropping, scaling, lossy compression, and combined attacks exploiting the properties of its perfect reconstruction, shift invariance, and good direction selectivity at the earliest. The experimental results demonstrate the superior performance than DWT-based watermarking schemes. Asikuzzaman et al.[11] combines SVD decomposition with DTCWT transform to construct a blind watermarking scheme making use of the stability of SVD singular values to improves the robustness of watermarking. Huan et al. [12] conducts the watermark embedding by modifying the candidate coefficients of joint sub-band of DTCWT, which is further improved over the previous scheme due to the modification is shared in joint sub-band and the main local energy of the joint sub-band is not significantly changed with the geometric attack.

Though these methods have strong robustness against signal attacks and geometric attacks, the robustness against temporal desychronization attack has not got enough attention. According to our experiments, most of the above-mentioned watermarking methods are not robust against frame rate conversion and frame swapping. It motivates us to propose a novel method to further improve the robustness to geometric and signal attacks while making the watermarking scheme resistant to temporal synchronization attacks.

### B. Dual Tree Complex Wavelet Transform

An important property of DTCWT is that its magnitude components are translation invariant. DCT divides the coefficients of video frames or video blocks into low frequency, medium frequency and high frequency. Since the lower-frequent part concentrates the main signal energy of the frame, modification towards it will greatly affect the perceptual quality of the video. At the same time, the higher-frequent part represents mostly the texture parts and contours of the video, which might not be robust against common attacks such as lossy compression and noise. Therefore, most DCT-based video watermarking schemes choose to modify the middle-frequent coefficients to achieve watermark embedding. As for transformation, the DWT transformation has disadvantages of lacking shift invariance and poor direction selectivity, and the CWT transformation has poor perfect reconstruction. DTCWT overcomes the disadvantages of both DWT and CWT by using two filter trees so that small changes in the input signal do not cause significant changes in the energy distribution of DTCWT coefficients at different scales. Besides, DTCWT is shift invariance, and has good direction selectivity and perfect reconstruction, which are not available in previous DCT, DFT, DWT and other transforms. Thus, the transformation is ideal for designing a video watermarking scheme that can resist geometric attacks such as rotation and scaling.

Applying 2D DTCWT to a video frame will produce six sub-bands corresponding to the output of each of the six angle-oriented directional filters. The complex wavelet coefficients at different levels can be defined as (1).

$$F_{l,d}(u_l, v_l) = |F_{l,d}(u_l, v_l)| e^{j\theta_{l,d}(u_l,v_l)} \quad (1)$$

where $d$ represents the direction of the filter, i.e., the six directional sub-bands. Fig.1 sketches the signal decomposition using four-level DTCWT. Each level includes a lower band as well as six higher sub-bands on different directions.

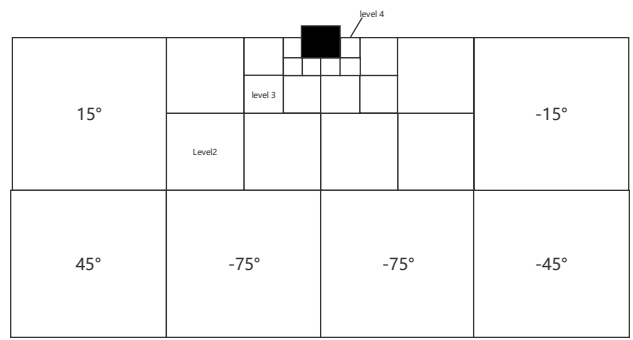

Fig. 1. Signal decomposition using four-level DTCWT. Each level includes a lower band as well as six higher sub-bands on different directions.

### C. Singular Value Decomposition

SVD decomposition is one of the highlights of linear algebra, which can diagonalize arbitrary matrices. A DTCWT sub-band is an array of nonnegative scalar terms, which can be regarded as a matrix. Performing SVD decomposition on DTCWT sub-bands makes the robustness of DTCWT-based watermarking methods further enhanced due to the good stability of singular values (SV). We denote the sub-band as a matrix $A$, and the singular value decomposition on $A$ can be performed as follows.

$$A = USV^H \quad (2)$$

where $U$ and $V$ are orthogonal matrices, $S$ is a diagonal matrix, and the descending diagonal elements of $S$ are called the singular values of $A$.

III. PROPOSED SCHEME

In this section, we propose the novel scheme that exploits temporal alignment of the watermark based on the duration of video to resist frame rate conversion.

A. Segmenting the video into clips

We segment the original video and watermarked video into clips. The choice of size of clip is a tradeoff between robustness and capacity. In particular, the larger size of clip, the smaller capacity and the stronger resistance to temporal synchronization attack. In contrast, the smaller size of clip, the larger capacity and the weaker resistance to temporal synchronization attack. In this paper, we propose a mechanism which can be capable to dynamically select the size of clip according to the number of frames and the duration of video clip size $c$ is calculated as $c = n \cdot d/c$, where $n$, $d$ and $c$ respectively denotes the number of frames, duration and a tradeoff parameter to balance the capacity and robustness. Note that for each segmentation, the total duration of the video is frame rate-invariant.

B. Candidate coefficient generation

Next, we divide the frames into the Y, U, V channels and select the U channel as embedding position due to its good perceptibility. We use three-level DTCWT transform to decompose U into one low-pass sub-band and six high-pass sub-bands. Each subband is associated with a different direction. Given a video with k frames, we denote the high-pass sub-band in the kth frame after DTCWT decomposition as $U_{l,d}^{H,k}$, where $l = 1, 2, 3$, $d = 1, 2, ...., 6$. We denote the low-pass sub-band as $U^L$, and we conduct SVD transformation on each high-pass sub-band respectively. We denote the singular value of every high-pass sub-band as candidate coefficients $SV_d^{i,k}$. The candidate coefficients are calculated by (4).

$$U_d^{i,k}(SV_d^{i,k})V_d^{i,k^T} = SVD(U_{3,d}^{H,k}) \quad (4)$$

C. Watermark Embedding Process

We denote the six candidate coefficients for each frame, which are $SV_1^{i,k}, SV_2^{i,k}, SV_3^{i,k}, SV_4^{i,k}, SV_5^{i,k}, SV_6^{i,k}$. The curve shape formed from six candidate coefficients before embedding a watermark generally is a "W"-shaped, with *Sub-band* 1 and *Sub-band* 4 at the bottom. Fig.2 shows the shape of curve formed from candidate coefficient of original video. We conduct the watermark embedding by adjusting the shape of the curve according to watermark. First, the curve formed from candidate coefficients of original frame is denoted as $p(d, y): y = p(d)$, where $d$ denotes the number of high-pass sub-band and $y$ denotes the corresponding singular value. Given a binary watermark bit $w$ whose value is either equal to 1 or -1, we conduct the watermark embedding by alter the overall trend of the curve. For simplicity, we remodel the curve as a monotonically increasing or monotonically decreasing line. We modify the candidate coefficient according to (5):

$$\begin{cases} \hat{p}: y = k(x - x_0) + y_0 & w = 1 \\ \hat{p}: y = -k(x - x_0) + y_0 & w = -1 \end{cases} \quad (5)$$

where $x_0$ and $y_0$ are the $x$ value and $y$ value of the points $p(0,0)$, $x$ value represents the number of high-pass sub-band and $y$ value represent the corresponding singular value, $k$ is a parameter greater than 0 indicating embedding strength. The magnitude of $k$ is a tradeoff between robustness and visual quality. Fig.3 shows the shape of curve formed from candidate coefficient of the watermarked video. It can be observed that the general trending of the curves is kept after performing the inverse DTCWT transform.

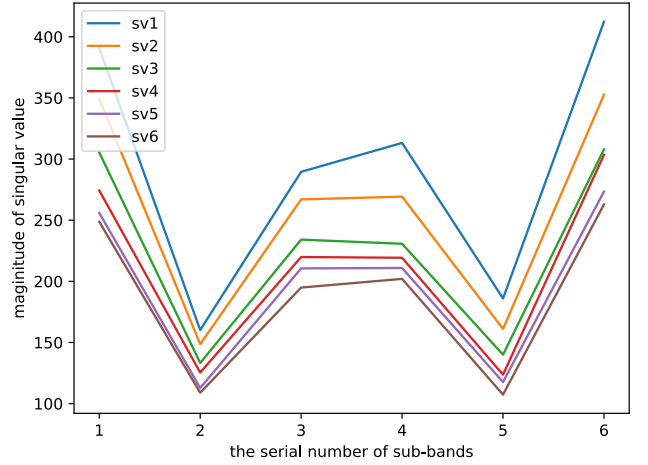

(a)

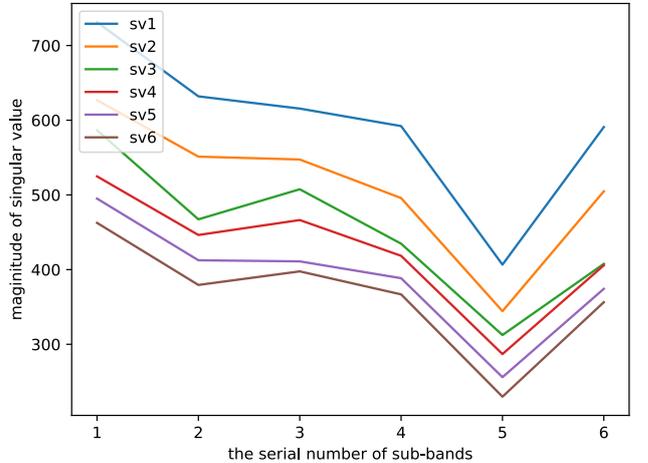

(b)

Fig. 2. Curves of averaged magnitude of singular decomposition values of the six candidate coefficients of (a) the original videos (b) the watermarked videos.

D. Watermark extracting process

Given a watermarked video, we obtain the candidate coefficient by performing SVD decomposition after DTCWT transform frame by frame. The binary bit embedded in the frame is extracted according to (6).

$$w = \begin{cases} -1, & if \quad SV_2^{i,k} < SV_4^{i,k} and SV_3^{i,k} < SV_5^{i,k} \\ 1, & otherwise \end{cases} \quad (6)$$

Then, we group the extracted binary sequences into group-level watermark according to the length of the video as described in Section III-A. We denote $flag = \sum_{i=1}^{i=c} w_i$ as standard for group-level watermark generation. The group-level watermark sequence is obtained by judging the relationship between $flag$ and zero. Finally, the group-level watermark is generated according to (7).

$$watermark_j = \begin{cases} 1, & if \quad flag > 0 \\ -1, & if \quad flag < 0 \\ 2, & otherwise \end{cases} \quad (7)$$

TABLE I. COMPARISON ON WATERMARKING IMPERCEPTIBILITY BETWEEN THE PROPOSED METHOD AND THE BASELINE METHODS

|  | Non-DTCWT[13] | DTCWT[11] | Proposed |
|---|---|---|---|
| PSNR(dB) | 36.14 | 38.05 | 39.52 |

TABLE II. COMPARISON ON ROBUSTNESS AGAINST VARIOUS KINDS OF ATTACKS.

|  | Normalized Correlation (NC) | | |
|---|---|---|---|
|  | Non-DTCWT[13] | DTCWT[11] | Proposed |
| Compression | 1.00 | 1.00 | 1.00 |
| Downscaling | 0.93 | 0.94 | 0.99 |
| Upscaling-cropping 110% | 0.83 | 0.96 | 0.99 |
| Upscaling-cropping 120% | 0.82 | 0.94 | 0.99 |
| Upscaling-cropping 130% | 0.81 | 0.95 | 0.98 |
| Rotation-cropping with 5 degrees | 0.89 | 0.97 | 0.99 |
| Rotation-cropping with 10 degrees | 0.74 | 0.96 | 0.99 |
| Rotation-cropping with 15 degrees | 0.57 | 0.96 | 0.99 |
| Frame rate 50fps | -- | -- | 0.99 |
| Frame rate 40fps | -- | -- | 0.99 |
| Frame rate 15fps | -- | -- | 0.99 |
| Frame rate 5fps | -- | -- | 0.99 |
| Rotation-cropping with 10 degrees-Frame rate 15fps | -- | -- | 0.99 |

## IV. EXPERIMENTS

### A. Experimental Setup

In order to verify the effectiveness of the duration of video-based watermarking scheme proposed in this paper, we use five HD videos ased from the Youtube-VOS dataset [23], including "Controlled_Burn", "ParkJoy", "SpeedBag", "Life", "Pedestrian_Area". We split the videos into twenty segmentations so that there are a total of 100 cover videos. We embed pseudo random binary bit sequence into video frame by frame which are grouped according to the duration of the cover video. $k$ is set as 0.8 for controlling the watermark strength, and $c$ is set as 1/6 for controlling the clip size. The experiments are conducted on a PC with Intel Core i7 processor.

### B. Performance evaluation

We evaluate the performance of our scheme in terms of robustness and visual quality. For evaluating the robustness we compute the NC (normalized correlation) value between original watermark and detected watermark. For evaluating the imperceptibility of watermarking embedding, we compute the PSNR (Peak Signal to Noise Ratio) value between original video and watermarked video.

**Visual quality**. Table 1 shows the comparison result between the proposed method and the three baseline schemes. As can be seen from the table, our scheme can preserve better visual quality than both non-DTCWT-based scheme [13] and DTCWT-based [11] scheme. Therefore, our scheme can ensure a better security and imperceptibility compared to previous works. The reason is that performing SVD decomposition on DTCWT sub-bands makes the robustness of DTCWT-based watermarking methods further enhanced due to the good stability of singular values.

**Robustness.** We compare the ability of the proposed scheme with a non-DTCWT-based state-of-the-art scheme [13] and a DTCWT-based state-of-the-art scheme [11] to extract the hidden information after the combined frame rate conversion and geometric attack. Table 2 shows the simulation results under various kinds of attacks. The experimental results show that the NC value of proposed method achieves lower than the other methods, and therefore the robustness of the algorithm is improved effectively. The results indicate that our scheme achieves superior robustness against both temporal and spatial attacks compared to previous methods.

## V. CONCLUSION

This paper proposes a novel and blind video watermarking scheme by embedding group-level watermarks into videos based on the DTCWT-SVD transformation. The secret information is hidden by grouping pseudo random binary bit sequences based on the duration of video. We embed bits by modifying the shape of the SVD curve of the candidate coefficient. Extensive experimental results show that the watermark can resist geometric attacks including cropping, rotation and scaling. Besides, the scheme is robust against frame rate conversion, which can satisfy both the requirements of blindly extraction and imperceptibility. In future work, we will further explore the relationship between DTCWT sub-bands to improve the capacity and performance of the scheme.

### ACKNOWLEDGMENT

This work is supported by National Natural Science Foundation of China under Grant U20B2051, U1936214. We thank the anonymous reviewers for their constructive and insightful suggestions on improving this paper. The authors are with the School of Computer Science, Fudan University, Shanghai, China. Email: {yfwang20, qcying20, zxqian, lisheng, zhangxinpeng}@fudan.edu.cn. Corresponding author: Zhenxing Qian.